\documentclass{emulateapj}
\usepackage{amsmath}
\usepackage{graphicx,natbib}
\usepackage{hyperref}
\newcommand{\sse}{\texttt{SSE}}
\newcommand{\mesa}{\texttt{MESA star}}
\newcommand{\bse}{\texttt{BSE}}
\shorttitle{Accelerated Stellar Evolution and Singleton sdB Stars}
\shortauthors{Clausen \& Wade}
\begin{document}
\title{How to Make a Singleton sdB Star via Accelerated Stellar Evolution}
\author{Drew Clausen \& Richard A. Wade}
\affil{Department of Astronomy \& Astrophysics, The Pennsylvania State University, 525 Davey Lab, University Park, PA 16802}
\email{dclausen@astro.psu.edu, wade@astro.psu.edu}

\begin{abstract}
  Many hot subdwarf B stars (sdBs) are in close binaries, and the favored
  formation channels for subdwarfs rely on mass transfer in a binary
  system to strip a core He burning star of its envelope.  However,
  these channels cannot account for sdBs that have been observed in long
  period binaries nor the narrow mass distribution of isolated (or
  ``singleton'') sdBs.  We propose a new formation channel involving the
  merger of a helium white dwarf and a low mass, hydrogen burning star,
  which addresses these issues.  Hierarchical triples whose inner
  binaries merge and form sdBs by this process could explain the
  observed long period subdwarf+main sequence binaries.  This process would also naturally explain the observed slow rotational speeds of singleton sdBs.  We also briefly
  discuss the implications of this formation channel for extreme
  horizontal branch morphology in globular clusters and the UV upturn in
  elliptical galaxies.
  \end{abstract}


\section{Introduction}   

Hot subdwarf B stars (sdBs) are thought to be core helium burning stars
with thin hydrogen envelopes \citep[for a recent review,
see][]{Heber:2009}.  Here, we define sdBs by observable quantities as
stars with $5.0 < \log g < 6.6$ and $20000 < T_{eff} < 45000$
\citep[see][]{Wade:2010}. To explain the origin of sdBs, a theory must
account for simultaneous mass loss and He ignition near the tip of the
red giant branch (RGB).  Many formation channels invoke binary mass
transfer to account for the loss of the H rich envelope
\citep[e.g.,][]{Mengel:1976,Han:2002,Han:2003}.  This mass loss
mechanism is supported by observations that show that 69\% of sdBs are
found in close binaries \citep{Maxted:2001}.  However, there are also
many ostensibly single sdBs, and \citet{Copperwheat:2011} presented a
revised estimate for the binary fraction in sdBs of only 51\%.  The
masses of some of these ``singleton'' sdBs have been estimated with
asteroseismology, and are seen to be narrowly distributed around
$0.47~M_{\sun}$\citep[e.g.,][]{Charpinet:1997, van-Grootel:2010}.  There
are many proposed formation channels for single sdBs, including the
merger of two He white dwarfs (WDs)
\citep{Webbink:1984,Iben:1984,Iben:1986}, enhanced RGB mass loss \citep{DCruz:1996},  ejection of the H
envelope by a sub-stellar companion \citep{Soker:1998}, and centrifugally enhanced mass
loss triggered by common envelope (CE) mergers \citep{Politano:2008}.  Recent observations of a sdB with a sub-stellar companion and a rapidly rotating, isolated sdB might be evidence of the latter two channels \citep[see][]{Geier:2011,Geier:2011a}, but otherwise, evidence
supporting the latter three channels is meager.   Furthermore, the population synthesis models presented by
\citet{Han:2002} suggest that WD mergers would lead to a wide
distribution in the masses of single sdBs, contrary to what is observed.
We propose that singleton sdBs can be the result of a binary merger, not
of two He WDs as previous studies have presented, but rather the merger
of a He WD and a very low mass hydrogen burning star.

\section{Accelerated Stellar Evolution}   

\begin{figure}[!]
\centering
\includegraphics[width=0.5\textwidth]{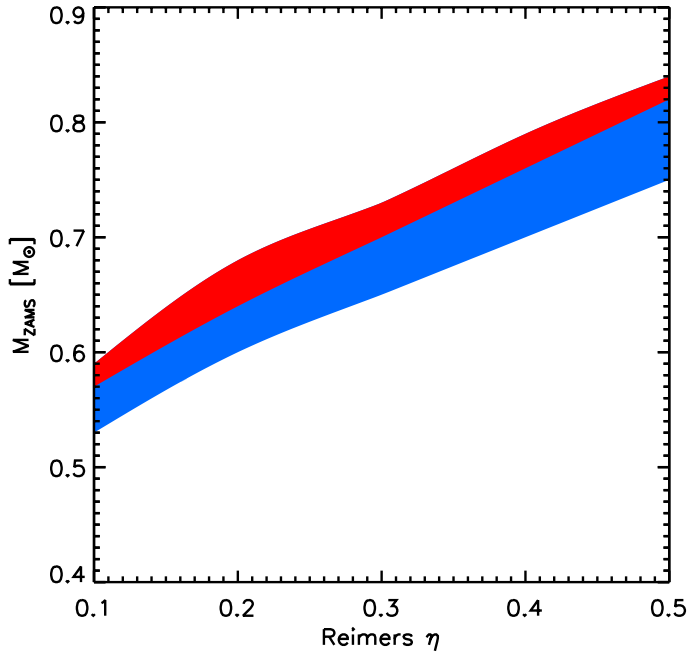}
\caption{ZAMS mass ranges that result in the formation of a sdB for different values of the Reimers mass loss parameter $\eta$.  For the upper, red region, we have assumed that He ignition occurs at the tip of the RGB.  If we allow stars to ignite He when the core has reached 95\% of the mass it will have at the tip of the RGB \citep{DCruz:1996}, the mass range includes the red region and extends to the lower values shown in blue.  These models were computed with \sse~assuming solar abundances.  This band shifts to the right for stars with $Y = 0.6$.   
  \label{fig:etaranges}}
\end{figure}

Given enough time, low mass stars can evolve directly to the sdB stage,
by way of RGB mass loss and He ignition under degenerate conditions.
Such low mass stars naturally lose their entire H envelopes by a
Reimers-like wind, and thus form singleton sdBs.  To delineate the ZAMS
mass range, RGB mass loss rate, and timescales required to form sdBs
from single stars, we ran stellar evolution models using the fast Single
Star Evolution code (\sse) described in \citet{Hurley:2000} and
confirmed the results with more detailed models using the
one-dimensional stellar evolution code \mesa~described in
\citet{Paxton:2011}.  The Reimers mass loss rate is given by $\dot{M} = 4\times 10^{-13}\; \eta R L/M~{\rm M_{\sun} yr^{-1}}$, where $R$, $L$, and $M$ are the star's radius, luminosity, and mass, respectively, and $\eta$ is a tunable parameter \citep{Kudritzki:1978}.  For values of $\eta$ in the range 0.1 - 0.5, stars with initial masses in the range
$M_{ZAMS} = 0.53 - 0.84~M_{\sun}$ will evolve to the sdB stage, see \autoref{fig:etaranges}.  The
resulting sdBs are all concentrated in mass at the usual He ignition mass at the tip of the RGB, $0.47 - 0.51~M_{\sun}$, which would account for the
observed narrow mass distribution of singleton sdBs with precise mass
determinations.  The problem is that the universe is not yet old enough
for this to have occurred in the single star context.  Given the usual
composition, $Y \sim 0.25-0.28$, it takes between 25 and 80 Gyr for
these single stars to evolve into sdBs.  However, if the evolution of
the star could be accelerated, singleton sdBs could form in the
observed mass range at the present epoch, directly from low mass stars.

\subsection{Binary Evolution}   

Injecting an already formed He core into a low mass star can ``create''
a $0.53 - 0.84~M_{\sun}$ star that is already at an advanced
evolutionary age, so that it can ascend the giant branch and ignite He
within a Hubble time.  The preformed He core is delivered in the form of
a He WD, the remnant of mass transfer in a previous stage of close
binary mass exchange.  This scenario requires a binary consisting of a
primary that is massive enough to have evolved to the RGB in less than a
Hubble time and an M dwarf companion.

We illustrate this formation pathway with an example computed using the
Binary Star Evolution code (\bse) described in \citet{Hurley:2002}.
Initially, the binary consisted of a $1.5 ~M_{\sun}$ primary and a
$0.25~M_{\sun}$ companion with an orbital period of 75 days.  After 2.9
Gyr, the primary began moving up the giant branch and expanded to
fill its Roche lobe, resulting in unstable mass transfer.  The system
went through a CE phase during which the primary's H
envelope was ejected.  The system emerged as a $0.33~M_{\sun}$ He WD and a
$0.25~M_{\sun}$ companion with an orbital period of 4 hours.  The system
then underwent a period of tidal readjustment that further reduced the
orbital separation and caused the low-mass, hydrogen burning star to
fill its Roche lobe.  Low mass main sequence stars are deeply
convective, so the mass transfer was unstable and the stars coalesced
into a $0.57~M_{\sun}$ star with a $0.33~M_{\sun}$ He core.  Using
\mesa~and \sse, we calculated how long it would take a $M_{ZAMS} =
0.6~M_{\sun}$ star to evolve to a similar structure and found times of
79 and 82 Gyr, respectively.  However, through accelerated stellar
evolution the coalesced binary reaches this evolutionary state in only
5.1 Gyr.  While \bse~was adequate for modeling the binary evolution, we
needed to use \sse~and \mesa~to investigate the evolution of the merger
product.

\subsection{Evolution of the Merger Product}   

\begin{figure}[!]
\centering
\includegraphics[width=0.5\textwidth]{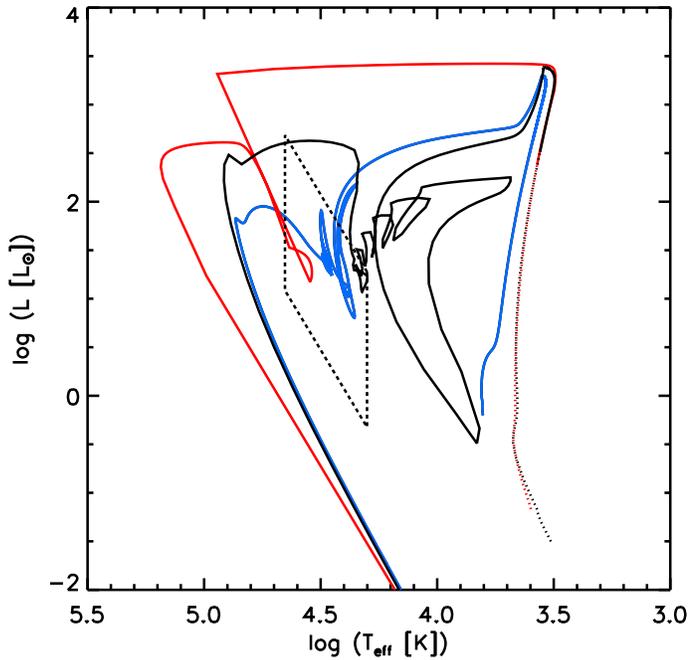}
\caption{$T_{eff}-L$ diagram showing three single stellar evolution models that
  result in the formation of a sdB.  The initial mass for each model is
  $0.6~M_{\sun}$.  The black and red curves show models with $Y = 0.28$
  and were computed with \mesa~and \sse, respectively.  The dotted
  portions of these curves are the $\sim 80$ Gyr of evolution that are
  ``skipped'' by merging a He WD with an M dwarf.  The excursion to low luminosity after reaching the tip of the RGB seen in the \mesa~model occurs on a timescale of $2\times10^{4}$ yr.  The blue curve shows
  a model with $Y = 0.6$ and was computed with \mesa.  The dashed lines show the ``sdB box'' for $0.48~M_{\sun}$ sdBs. 
  \label{fig:tracks}}
\end{figure}

After the binary coalesced, \bse~continued to evolve the merger product
as a single star and, in the illustrative case described
above, the merger product eventually reached the sdB stage.  However,
based on the merger product's core mass,~\bse~assumed that it was a
$M_{ZAMS} = 2.5~M_{\sun}$ star at the base of the RGB and evolved the
star accordingly.  This assumption is clearly not appropriate to model
the evolution of a $0.57~M_{\sun}$ star, so we used \sse~and \mesa~ to
find stars with core-envelope structures similar to the merged star from
the \bse~model.  As described above, the stellar evolution models showed
that a star with $M_{ZAMS} = 0.6~M_{\sun}$ eventually formed a
$0.33~M_{\sun}$ He core surrounded by a $\sim 0.25~M_{\sun}$ H rich
envelope.  When we continued the evolution of this star, the mass of the
He core grew through H shell burning and ignited at the tip of the giant
branch while Reimers mass loss removed the remaining H envelope.  This
additional evolution from the RGB to the sdB stage took 140 Myr.  The
evolutionary tracks computed by \sse~and \mesa~are shown in
\autoref{fig:tracks}.  If the merged He WD--M dwarf maintains the
distinct core-envelope structure of an RGB star, then the binary
evolution scenario described above forms a $0.48~M_{\sun}$ sdB in 5.2
Gyr.

\begin{figure*}[!]
\includegraphics[width=\textwidth]{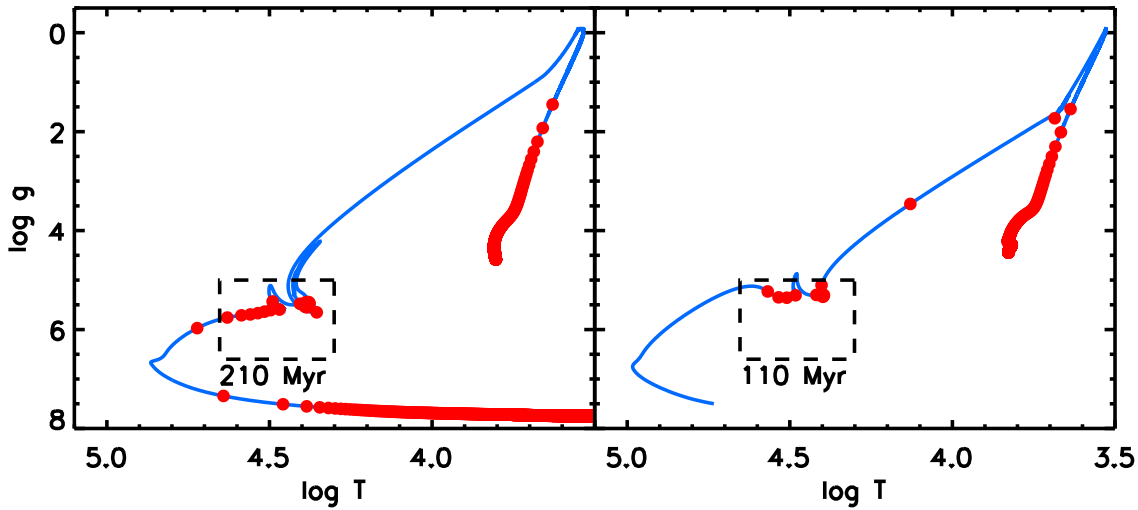}
\caption{Surface gravity vs. effective temperature for $Y = 0.6$ stars.
  The left and right panels show the evolution of the quantities for
  stars of initial mass $0.6~M_{\sun}$ and $0.7~M_{\sun}$, respectively.
  The dots are evenly spaced in time, each interval corresponding to 10 Myr
  of evolution.  Dashed lines show the ``sdB box'' and the total
  duration of the sdB phase for each star is noted on the
  plot.\label{fig:logtg}}
\end{figure*}

It is not clear whether the merged star will maintain the
core-envelope structure of an RGB star.  Instead, some mixing might occur that would alter the chemical profile assumed above.  To bracket the range of possible outcomes, we explored models in which the He WD mixed completely 
with the M dwarf and formed a homogeneous, He-rich star.  In the mergers
considered here, we are mixing $\sim 0.2~M_{\sun}$ of material with
standard abundances with $\sim 0.3~M_{\sun}$ H depleted material,
resulting in a star with $Y \sim 0.6$.  Using \mesa, we have modeled the
evolution of these completely mixed stars and found that they too will form sdBs.
\autoref{fig:logtg} shows evolutionary tracks in the ($\log~T_{eff},
\log~g$) plane for $0.6~M_{\sun}$ and $0.7~M_{\sun}$ mixed stars with $\eta = 0.5$ and 0.7, respectively, and
initial $Y=0.6$.  The less massive star takes 5.8 Gyr to become a sdB
and remains in the sdB ``box'' for 210 Myr, while the more massive star
evolves to the sdB stage in only 3.3 Gyr and remains in this stage for
110 Myr.  The sdB stars have masses of $0.44~M_{\sun}$ and
$0.50~M_{\sun}$, respectively.  For comparison, the evolutionary track of
the $0.6~M_{\sun}$ mixed star is also shown in \autoref{fig:tracks}.
These models demonstrate that even if the preformed He core dissolves
during the merger, these systems can still evolve into sdBs within a
Hubble time.

\section{Discussion}   

This formation channel depends on several assumptions made in modeling
binary stellar evolution.  The initial configuration of the binary must
be such that the primary evolves to the giant branch and triggers an
episode of unstable mass transfer that completely removes its envelope
before the merger.  Furthermore, the total mass of the merged star must
be in the range $0.53 - 0.84~M_{\sun}$ for the merged object to evolve
into a sdB.  We have computed a grid of models with $\eta = 0.3$ to make a preliminary
exploration of the parameter space of progenitor systems with primary
masses in the range $1 - 3~M_{\sun}$, secondary masses in the range $0.1
- 0.8~M_{\sun}$, periods in the range $1 - 350$d, and CE
ejection efficiencies, $\alpha_{CE}$, in the range $0.5 - 1.5$.  Of the
$5\times10^{4}$ models in our grid, 6\% produce merger products that
will evolve into sdBs within a Hubble time.  We have assumed that the core-envelope structure is retained and used the results of \autoref{fig:etaranges}. The primary masses in these
systems range from $1.2 - 3~M_{\sun}$, the secondary masses range from
$0.1 - 0.6~M_{\sun}$, and the initial periods range from $10 - 350$d.  In many cases, the CE phase was sufficient to remove the primary's envelope and drive the system to merger so that the tidal readjustment phase described above was not required.  
For each value of $\alpha_{CE}$ roughly the same number of proto-sdBs
were formed, except for $\alpha_{CE} = 0.5$ which produced 30\% fewer
sdB progenitors.  In this case many binaries merge before the primary's
envelope is removed, producing an RGB star that is too massive to evolve
directly to the sdB stage.  A full exploration of the parameter
space and formation rate requires further work, but we note that our preliminary
investigation suggests that a diverse population of initial binaries
will evolve into singleton sdBs and that this result holds for a wide
range of values for $\alpha_{CE} $ and $\eta$.  Furthermore, since both
members of the initial binary are of relatively low mass, their
formation is favored by the observed Initial Mass Function.

The time required to form a singleton sdB with this channel varies
widely.  In one extreme, a binary consisting of stars with masses of
$3~M_{\sun}$ and $0.35~M_{\sun}$ with a 90 d period merged after only
380 Myr, implying that if the merger product maintains its core-envelope
structure, this system could form an sdB within $\sim 0.5$ Gyr of its
birth.  On the other hand, some systems take more than a Hubble time to
coalesce and, if the merged star mixes it could take an additional 6 Gyr
to evolve to the sdB stage.  From our grid of models, the mean amount of
time for a system to merge into a proto-sdB was 5.5 Gyr.  More work is
needed to study the chemical stratification of the merger product, but
the time it takes the merged star to become an sdB is bracketed by the
140 Myr and 3-5 Gyr time scales for the non-mixed and completely mixed
cases, respectively.
\subsection{Long Period sdB+Main Sequence Binaries} 
This channel may also explain a conundrum among the presently observed
sdB + G or K dwarf binaries.  
(We will use ``MS'' as shorthand notation for G and K dwarfs.)
\citet{Han:2003} predicted that all such
systems form as the result of a CE phase and should have
periods $\la 20$ d.  These authors also predict long period $(P\ga40$ d), post-Roche lobe overflow sdB + G or K binaries, but in these systems the companions are subgiants or giants (i.e., more massive stars at a later evolutionary state). The short period, sdB+MS binaries should be easy to find
because their large velocity variations can easily be discerned within a
single observing run, but none have been reported.  The observational evidence suggests that the presence of a G or K dwarf companion indicates a wide ($P>100$~d) binary \citep[see, e.g.,][and references therein]{Copperwheat:2011}, despite the suggestion of \citet{Heber:2002} that radial velocity observations should reveal such sdB+MS systems to be close.  Our own experiments with \bse, including various
modifications to the mass loss, angular momentum loss, and stable mass
transfer criterion \citep[some of which mimic the results of][]{Han:2003},
fail to produce long period sdB+MS binaries. But if these sdB+MS binaries are instead viewed as the binary remnants of original
hierarchical triple systems, in which the inner binary has evolved to
become a singleton sdB as outlined above, then the remaining outer
binary (presently seen as sdB+MS) was never ``close'' (i.e., tidally
interacting) and is thus irrelevant to the production of the sdB.

For stability of the hierarchical triple, the ratio of the semi-major
axis of the outer binary to that of the inner, sdB forming binary must
be greater than $\sim 20 \log(1 + m_{3}/m_{B})$, where $m_{3}$ is the
mass of the outer star, $m_{B}$ is the mass of the inner binary, and we
have assumed circular orbits \citep{Harrington:1975}.  Furthermore, as
mass is lost by the inner binary to form the sdB, the orbit of the outer
binary will expand adiabatically to $a_{f} = a_{i} (M_{i}/M_{f})$,
where $a$ is the semi-major axis and $M$ is the total mass of the system
and the subscripts $i$ and $f$ correspond the value before and after
mass loss, respectively \citep{Eggleton:1989, Debes:2002}.  If we apply these
constraints to the illustrative case described above and assume that
this binary is orbited by a $0.8~M_{\sun}$ K dwarf, the {\it minimum}
orbital period of the resulting sdB + K dwarf binary is 1360 d.  When we
consider the entire grid of models discussed above, the shortest
possible period for a sdB + 0.8 $M_{\sun}$ K dwarf binary is 185 d.
Furthermore, we note that the outer star might promote the merger of the inner binary via the Kozai mechanism.   This triple-star channel, involving the new H-merger channel
described above, can produce long period sdB+MS binaries,
so previous studies of the sdB+MS binary population that
do not include this channel are incomplete.

\newpage

\subsection{Rotation of Single sdBs}
\citet{Gourgouliatos:2006} studied the merger of two He WDs assuming complete conservation of momentum and found that the rotational velocity of the merger product could be $\ga 10^{3} {\rm~km~s^{-1}}$.  While there is undoubtedly some angular momentum lost during the merger, it is difficult to explain why nearly all single sdBs observed have projected rotational velocities of $< 10 {\rm~km~s^{-1}}$ \citep{Geier:2009a}.  An advantage of the new H-merger channel described above is that the merger product will lose between $\sim 0.1-0.3~M_{\sun}$ of material while it is on the giant branch.   Angular momentum carried away by this material can spin down the star, resulting in a slowly rotating sdB.    

\subsection{Other Implications}  

Contributions from the He WD + M dwarf formation channel for sdBs
presented here might also play a role in determining the binary fraction
of extreme horizontal branch (EHB) stars in globular clusters, and in the
UV-upturn in elliptical galaxies.  The short-period binary fraction among EHB stars
in globular clusters is much lower than that of field sdBs, and this has
been attributed to the fact that the dominant formation channel for EHBs
in old stellar populations is He WD mergers that result in a singleton
sdB \citep{Moni-Bidin:2008,Han:2008,Moni-Bidin:2011}.
The formation channel presented here also produces singletons, and in
some cases, especially if the merger product is mixed as considered above,
the process can take $\ga 12$ Gyr to produce a sdB. These mixed stars would exhibit the super-solar helium abundance invoked in some EHB models \citep[e.g.,][]{Sweigart:1979,Dalessandro:2011},
although star-by-star rather than as a population.  Again, more work is
needed to determine whether this channel contributes to the EHB
population significantly at late times.

Finally, we note that sdBs produced by this merger channel could also
contribute to the UV-upturn observed in elliptical galaxies.  The
evolved stellar populations thought to inhabit elliptical galaxies would
not produce the UV-excess seen in their spectral energy distrubutions, and \citet{Han:2007}
proposed that emission from the sdBs formed through binary evolution
might be the source of this radiation.  The formation channel presented here offers an
additional population of sdBs that supplies UV photons, perhaps on a
different timescale.  Determining the contribution of singleton sdBs
formed by He WD + M dwarf mergers to either the globular cluster or
elliptical galaxy populations is further complicated by metallicity effects.
\section{Conclusions} 

We have shown that merging a He WD with an M dwarf can produce a low
mass star of advanced evolutionary age or a helium rich star, either of
which can evolve to become a sdB within a Hubble time.  This model can
explain the narrow mass range in singleton sdBs and the existence of
long period sdB+MS binaries, if these systems
were initially triples. The sdBs produced by this formation channel
might also contribute to the low binary fraction among EHB stars in
globular clusters and the UV-excess in elliptical galaxies.  Many
aspects of this channel remain to be explored, including the formation
rate, the effect of metallicity variations, and the exact chemical
profile of the merger product.  We offer it as a supplementary and
possibly dominant channel for formation of singleton sdBs.

\acknowledgments

We thank the anonymous referee for insightful comments that improved the manuscript.  We are grateful to Steinn Sigurdsson and Mike Eracleous for helpful discussions.  Partial support for
this work was provided by the National Aeronautics Space Administration
(NASA) through Chandra Award Number TM8-9007X issued by the Chandra
X-ray Observatory Center, which is operated by the Smithsonian
Astrophysical Observatory for and on behalf of NASA under contract
NAS8-03060.  Support from NASA grant NNX09AC83G and National Science
Foundation grant AST-0908642 is also gratefully acknowledged.

\end{document}